\documentclass[conference]{IEEEtran}
\IEEEoverridecommandlockouts
\usepackage{cite}
\usepackage{amsmath,amssymb,amsfonts}
\usepackage{graphicx}
\usepackage{textcomp}
\usepackage{xcolor}
\usepackage{bm}
\usepackage{subfigure}
\usepackage{algorithm}
\usepackage{algpseudocode}
\usepackage{algorithmicx}
\usepackage{booktabs}
\usepackage{amssymb}
\def\BibTeX{{\rm B\kern-.05em{\sc i\kern-.025em b}\kern-.08em
		T\kern-.1667em\lower.7ex\hbox{E}\kern-.125emX}}
\begin{document}
	
	\title{CKMImageNet: A Comprehensive Dataset to Enable Channel Knowledge Map Construction \\via Computer Vision }
	
	\author{\IEEEauthorblockN{$\text{Di~Wu}^*$, $\text{Zijian~Wu}^*$, $\text{Yuelong~Qiu}^*$, $\text{Shen~Fu}^*$, $\text{Yong~Zeng}^*\ddagger$, }
		\IEEEauthorblockA{*National Mobile Communications Research Laboratory, Southeast University, Nanjing 210096, China\\
			$\ddagger$Purple Mountain Laboratories, Nanjing 211111, China\\
			studywudi@seu.edu.cn, 220231172@seu.edu.cn, yl\_qiu@seu.edu.cn, 230248151@seu.edu.cn, yong\_zeng@seu.edu.cn}
	}
	
	\maketitle
	
\begin{abstract}
Environment-aware communication and sensing is one of the promising paradigm shifts towards 6G, which fully leverages prior information of the local wireless environment to optimize network performance.
One of the key enablers for environment-aware communication and sensing is channel knowledge map (CKM), which provides location-specific channel knowledge that is crucial for channel state information (CSI) acquisition.
To support the efficient construction of CKM, large-scale location-specific channel data is essential. However, most existing channel datasets do not have the location information nor visual representations of channel data, making them inadequate for exploring the intrinsic relationship between the channel knowledge and the local environment, nor for applying advanced artificial intelligence (AI) algorithms such as computer vision (CV) for CKM construction.
To address such issues, in this paper, a large-scale dataset named CKMImageNet is established, which can provide both location-tagged numerical channel data and visual images, providing a holistic view of the channel and  environment.
Built using commercial ray tracing software, CKMImageNet captures electromagnetic wave propagation in different scenarios, revealing the relationships between location, environment and channel knowledge.
By integrating detailed channel data and the corresponding image, CKMImageNet not only supports the verification of various communication and sensing algorithms, but also enables CKM construction with CV algorithms.

\end{abstract}
	
\section{Introduction}

The evolution of wireless communication technologies has driven an unprecedented demand for high data rates, low latency, and enhanced reliability. As the transition towards the sixth-generation (6G) wireless network progresses, the objective is to achieve significant performance improvements through the integration of innovative technologies, such as millimeter-wave (mmWave), extremely large-scale multiple-input multiple-output (XL-MIMO), Terahertz communications, and advanced modulation techniques \cite{3021,3025,3022,3029,lu2024tutorial}. 
These advancements promise substantial enhancements in data throughput and spectral efficiency, setting new benchmarks for future wireless networks. However, these ambitious visions bring forth significant challenges. In particular, the increasing density of network nodes and the higher dimensionality of communication channels result in considerable overhead for channel state information (CSI) acquisition. Furthermore, 6G networks demand precise synchronization and coordination among numerous devices, increasing the requirements for real-time and global CSI. On the other hand, 6G also offers many new opportunities to address such challenges, including the proliferation of devices and the integration of advanced technologies that enable the collection of large-scale communication and environment data, enhanced localization capabilities for precise device tracking, and integrated localization, sensing and communication (ILSAC) \cite{2102,xiao2022overview}. These opportunities catalyze a paradigm shift from the conventional environment-unaware towards environment-aware communication and sensing \cite{2101,zeng2024tutorial}, which leverages detailed insights into the wireless environment to enhance CSI acquisition  and optimize network performance.

A key enabler of environment-aware communication and sensing is channel knowledge map (CKM), which provides location-specific channel knowledge useful to enhance environment-awareness and facilitate or even obviate sophisticated real-time CSI acquisition \cite{2101,zeng2024tutorial}. CKM reveals intrinsic relationships between location and wireless channel, which is crucial for network optimization. Built on CKM, channel twins offer virtual replicas of physical communication channels, created using detailed channel data and simulations. These digital twins enable real-time analysis and adaptive communication strategies by replicating the physical environment's behavior and interactions \cite{jian2021study,wang2024digital}. Furthermore, artificial intelligence (AI) may play a pivotal role in environment-aware communication by leveraging large-scale channel data to develop predictive models and optimization algorithms \cite{letaief2019roadmap,yang2020artificial}. Advanced AI techniques may enhance CKM construction, improve map resolution and accuracy, and enable real-time analysis and adaptive strategies for channel twins. 
In particular, computer vision (CV) is appealing for such tasks, which excel in processing data that is represented as graphs, making them ideal for constructing CKM. CV facilitates the efficient learning and prediction of channel features, improving CKM resolution and accuracy \cite{voulodimos2018deep,szeliski2022computer}. They also enhance the real-time analysis and adaptive strategies for channel twins, enabling more effective and responsive network management.

The successful implementation of these environment-aware technologies critically depends on large-scale channel datasets. 
Many channel datasets have been established to support AI training, validate algorithm performance, and facilitate various research and development activities in wireless communications\cite{alkhateeb2019deepmimo,shen2023dataai,klautau20185g,cheng2023m}. These datasets are instrumental in developing and testing new communication protocols and optimization techniques. However, existing channel datasets have several limitations. Firstly, they usually lack a clear relationship between channel data and user locations, making them impossible to support the construction of  CKM. Additionally, most datasets do not incorporate visual representations of channel characteristics, such as channel heatmaps or physical environment maps, which are crucial for CV-based CKM construction algorithms. The absence of (relative) location information and visual data severely hinders the exploration of intrinsic relationships between channel and environmental factors, limiting the effectiveness of AI-driven applications and reducing the overall utility of these datasets for advanced environment-aware communication research.

\begin{figure}[htbp]
	\centering{\includegraphics[width=.48\textwidth]{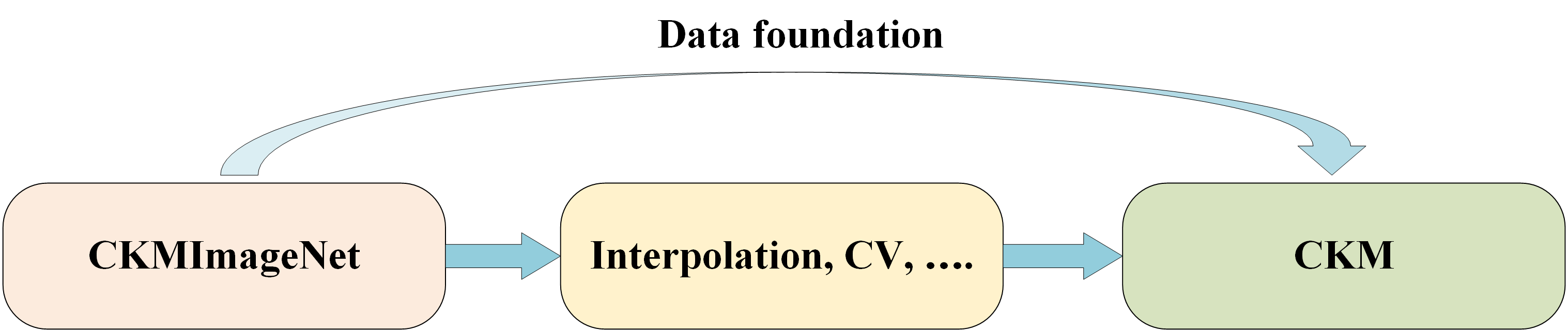}}  	
	\caption{The relationship between CKMImageNet and CKM. } 
	\label{CKM}  
\end{figure}

To address these limitations, in this paper, a mixed data and image dataset is constructed, named CKMImageNet\footnote{The CKMImageNet can be found on the website \cite{ckmimagenet}}, intending to provide a detailed location-based channel knowledge across various scenarios.
As shown in Fig. \ref{CKM}, CKMImageNet provides the data foundation for CKM, facilitating the design and validation of CKM construction algorithms, such as data interpolation and CV. 

\section{CKMImageNet: Features and Framework}
In this section, we explain the unique features and the comprehensive framework of the developed CKMImageNet. CKMImageNet is established to offer channel data in form of both numerical and visual data to support advanced wireless communication research and CKM construction.

\subsection{Features of CKMImageNet}
As shown in Table \ref{tablecompare}, the CKMImageNet have several distinctive features that set it apart from conventional channel datasets:
\begin{itemize}
	\item \textbf{Data with location information:} A key feature of CKMImageNet is its ability to associate channel data with the locations of base stations (BSs) and user equipment (UEs). This spatial correlation is crucial for accurate modeling and construction of CKM, as it enables a detailed understanding of how the channel characteristics vary across different locations.
	\item \textbf{Comprehensive channel knowledge data:}
	CKMImageNet includes an extensive range of channel parameters such as path loss, delay spread, angle of arrival (AoA), and angle of departure (AoD). These parameters are critical for a thorough understanding of signal propagation and are essential for the design and optimization of advanced communication systems. By providing detailed channel data, the CKMImageNet enables the construction of diverse types of CKM and supports a wide range of application scenarios.
	\item \textbf{Visual CKM:}
	CKMImageNet offers a range of visual CKM, which provides an intuitive understanding of channel characteristics, making it easier for users to interpret and utilize the data. Visualizations help in identifying patterns and understanding spatial relationships, thereby aiding in CV training.
	\item \textbf{Environment information integration:}
	CKMImageNet incorporates comprehensive information about the physical environment, including the physical environment map and materials of buildings, terrain features, and other relevant structures. By providing not only channel knowledge but also physical environment knowledge, CKMImageNet facilitates the development of advanced CKM construction and inference algorithms that fuse both channel knowledge data and environment data.

	\item \textbf{Multi-scenario support:} CKMImageNet is designed to support a wide variety of scenarios, encompassing urban, rural, and indoor environments. This multi-scenario support allows researchers and engineers to analyze and optimize wireless networks across different contexts. The ability to simulate and study various environments makes the CKMImageNet a versatile tool, useful for a broad range of applications and research projects.
	\item \textbf{High fidelity through ray tracing:}
	CKMImageNet utilizes advanced ray tracing techniques to simulate electromagnetic wave propagation with high fidelity by using the commercial ray-tracing software Wireless Insite \cite{wirelessinsite}. Ray tracing models the interactions of waves with the environment, including reflections, diffractions, and scatterings. This high level of fidelity ensures that the generated channel data is reliable, providing a solid foundation for advanced network design and analysis.
	\item \textbf{Customization for application requirement:}
	The CKMImageNet is highly customizable to meet specific application requirements. CKMImageNet focuses on the intrinsic characteristics of the channel, independent of transmitter and receiver configurations, which means users can tailor the dataset to their particular requirements by configuring the transmitter and receiver settings and parameters. This flexibility ensures that the CKMImageNet can be effectively used for targeted research and practical applications, making it a valuable resource for a wide range of projects and studies.

\end{itemize}

In summary, CKMImageNet stands out with its data and location correlation, comprehensive channel knowledge data, visual CKM, support for multiple scenarios, high fidelity, and customization for specific application requirements. These features collectively make the CKMImageNet an invaluable tool for advanced wireless communication research and AI-based applications, providing a robust data foundation for both current and future technological developments in the field.

\begin{table*}[htbp]
	\centering
	\caption{Comparison between CKMImageNet and existing datasets}
	\begin{tabular}{ccccc}
		\toprule  
		  & CKMImageNet&DeepMIMO\cite{alkhateeb2019deepmimo}&DataAI-6G\cite{shen2023dataai}&M$ ^3 $SC\cite{cheng2023m} \\ 
		\midrule  
		Data with location information &\checkmark&$ \times $&$ \times $&$ \times $ \\
	    Detailed channel knowledge data&\checkmark&\checkmark&\checkmark&\checkmark \\
		Visual CKM & \checkmark&$ \times $&$ \times $&$ \times $\\
		Environment information integration & \checkmark &$ \times $&$ \times $&\checkmark\\
		Multi-scenario support & \checkmark&$ \times $&$ \times $&$ \times $\\
		Customization for application requirement & \checkmark&\checkmark&\checkmark&\checkmark\\
		\bottomrule  
	\end{tabular}
\label{tablecompare}
\end{table*}

\begin{figure*}[htbp]
	\centering{\includegraphics[width=0.98\textwidth]{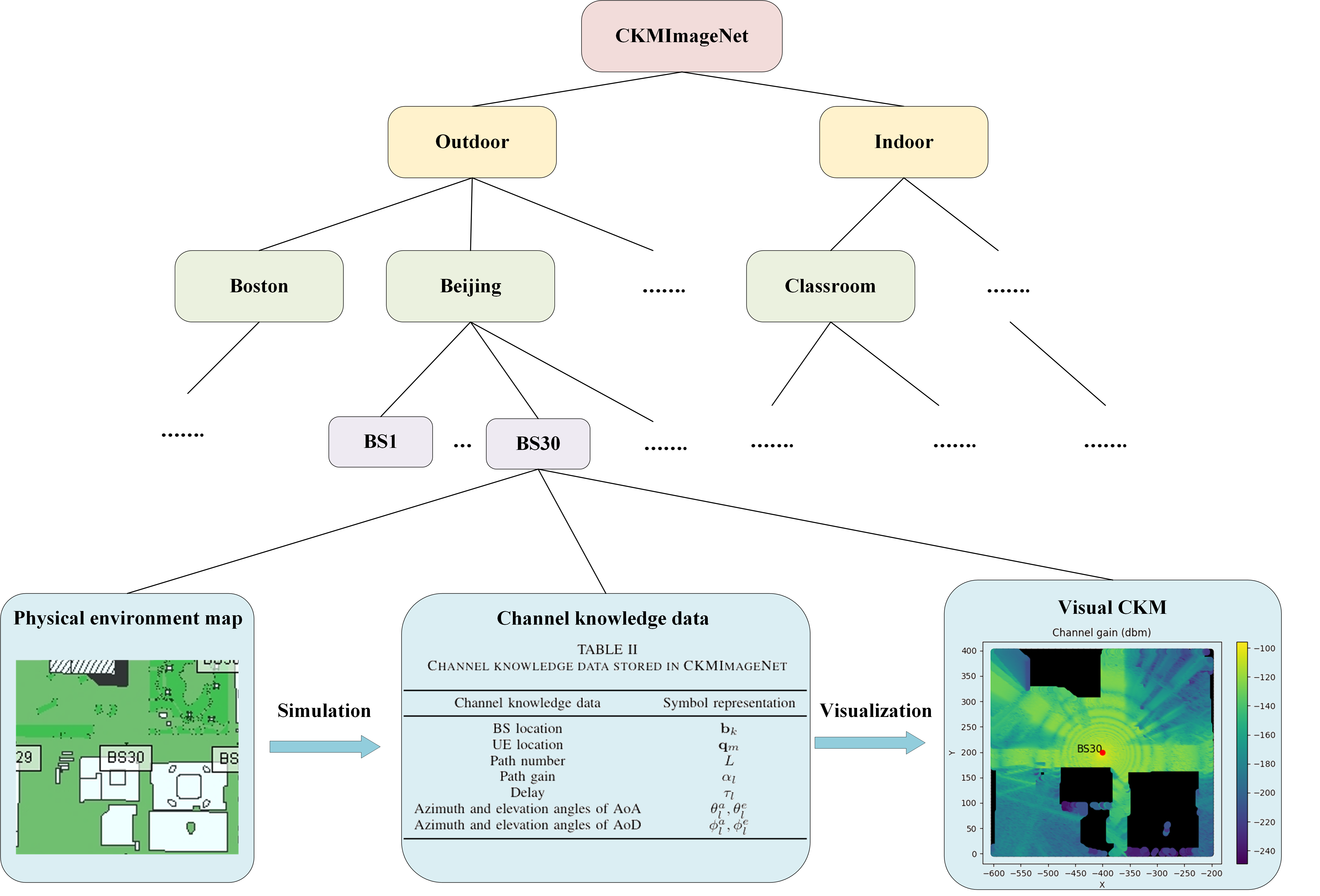}}  	
	\caption{The framework of CKMImageNet. } 
	\label{framework}  
\end{figure*}

\subsection{Framework of CKMImageNet}
As shown in Fig. \ref{framework}, the CKMImageNet consolidates all data into a net-like hierarchical structure, organized from top to bottom as scenarios, BS, and data. Users can step by step select the required data according to this hierarchical structure. The final selectable data includes the following components:
\begin{itemize}
\item \textbf{Physical environment map:}
The foundation of the CKMImageNet framework is the environment map, which provides a detailed representation of the physical environment. The environment maps in CKMImageNet are digitized from actual maps, including the geometry of buildings, terrain features, and other relevant structures. 
At present, CKMImageNet has integrated up to 10 different types of environment maps, including various application scenarios such as urban, rural, indoor, and outdoor settings.
By extracting from these scenarios, more than 10,000 different environment maps, each with size 128 meter x 128 meter, have been obtained and stored in CKMImageNet.
Alongside the map, the dataset stores detailed environment data, such as the materials of surfaces, which influence signal reflection, diffraction, and scattering. This comprehensive environment context ensures that the channel data accurately reflects real-world conditions, enables users to directly visualize the signal propagation environment and explore the relationship between the channel and environment.

\item \textbf{Channel knowledge data:}
The CKMImageNet includes extensive channel data. At present, over 25 million channel data entries have been stored. 
Each data entry includes information about the BS locations $ \mathbf{b}_k, k=1,..,K $, UE locations $ \mathbf{q}_m, m=1,..,M $ and channel parameters of each path $ l, l=1,...,L $, including path gain $ \alpha_l $, delay $ \tau_l $,  azimuth and elevation angles of AoA $ \theta_l^a, \theta_l^e$, and azimuth and elevation angles of AoD $ \phi_l^a, \phi_l^e$, which is shown in Table II in Fig, \ref{framework}.


This data is generated through ray tracing simulations, which model the interactions of electromagnetic waves with the environment. 
The channel data is stored in a net-like structured format, allowing for efficient querying and retrieval.

\item \textbf{Visual CKM:}
To facilitate CV algorithms, the CKMImageNet includes visual representations of the channel data in the form of visual CKM, which display the spatial distribution of key parameters such as signal strength and channel angle distribution across the environment. 
As of now, CKMImageNet has stored 40,000 channel knowledge images, each sized 64$ \times $64 pixels, comprising 4,096 pixels corresponding to channel knowledge at different locations.
These visualizations help users to quickly identify patterns and areas of interest, making it easier to diagnose network performance issues and plan optimizations. This visual component of the CKMImageNet enhances its usability, providing an accessible way to comprehend complex channel characteristics, which is suitable for use as a training dataset for CV.

\end{itemize}

In summary, the CKMImageNet framework combines detailed physical environmental maps, extensive channel data, and intuitive visualizations to provide a comprehensive and accessible resource for wireless communication research and development. 
As shown in Fig. \ref{framework}, these three data types are interdependent: environment map forms the basis for generating channel data. Channel data can be further visualized to create channel heatmaps. Channel heatmaps, in turn, provide an intuitive representation of the relationship between the channel and the environment.
By organizing the data in a net-like structured and efficient manner, the CKMImageNet ensures that users can easily access and leverage the information to design, optimize, and analyze advanced communication systems.

\section{Constructing CKMImageNet}
To illustrate the process of constructing CKMImageNet, we take the urban environment of Beijing as an example. This section details each step involved in building a comprehensive CKMImageNet, including scenario construction, BS and UE setup, the ray tracing process, and the structure of both channel data and channel heatmaps.

\begin{figure}[htbp]
	\centering{\includegraphics[width=.5\textwidth]{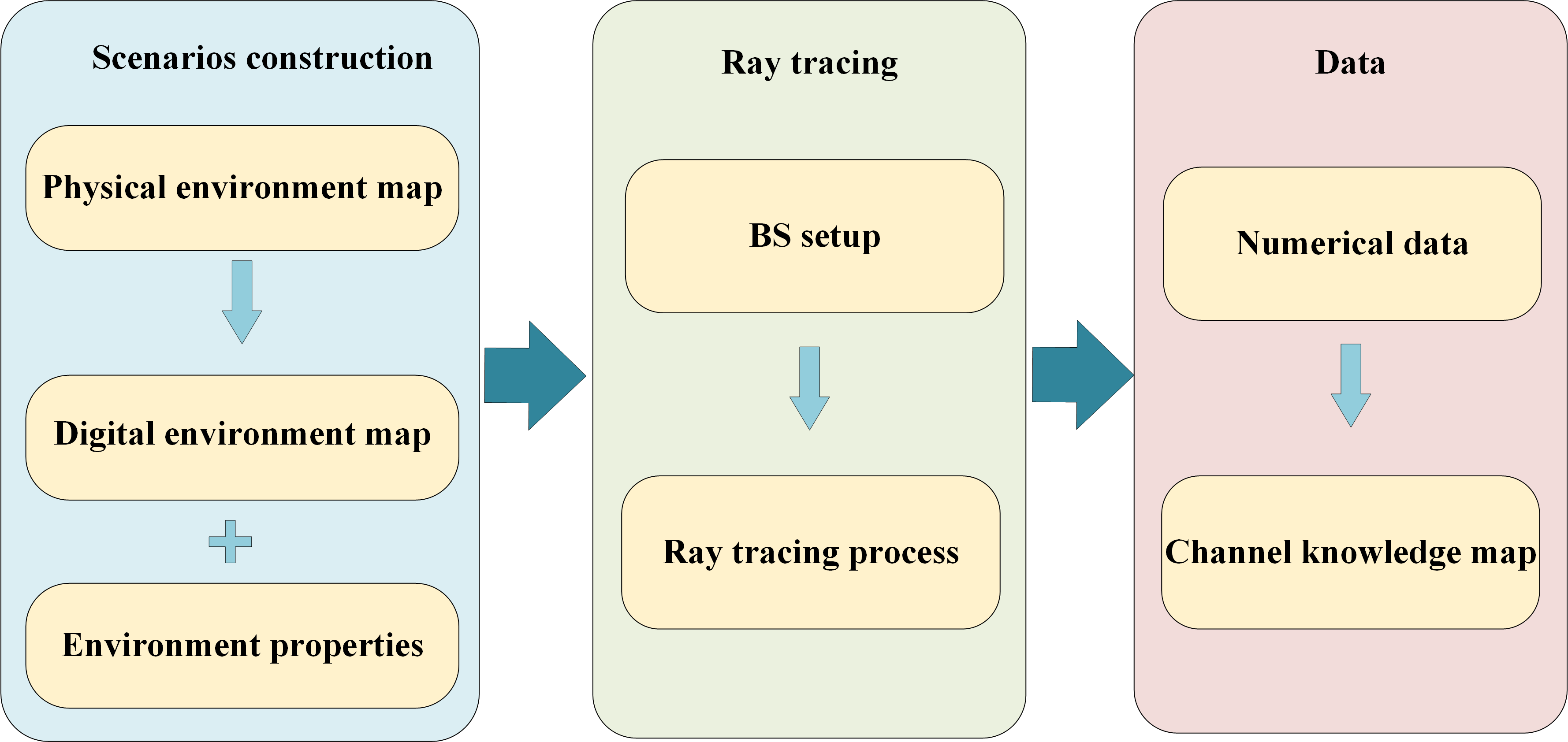}}  	
	\caption{The construction process of CKMImageNet. } 
	\label{construction}  
\end{figure}

\subsection{Scenarios construction}
In constructing the CKMImageNet for Beijing, the first step is to accurately model the city's environment. This involves capturing detailed information about the city's physical features and creating a realistic simulation environment. 

\begin{figure}[htbp]
	\centering{\includegraphics[width=.4\textwidth]{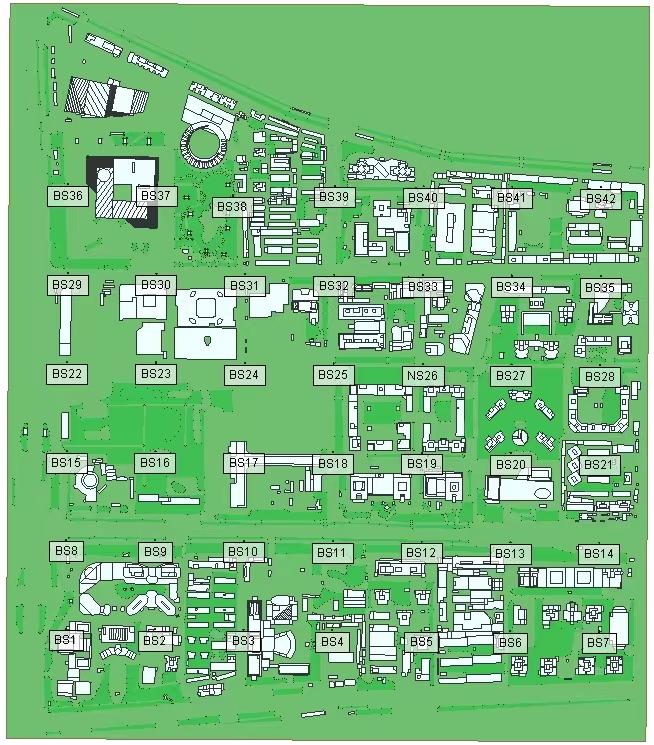}}  	
	\caption{The environment map of a particular region of Beijing. } 
	\label{Beijing}  
\end{figure}

\begin{itemize}
\item \textbf{Urban environment modeling:} The city of Beijing is characterized by a mix of high-rise buildings, residential areas, commercial districts, and open spaces. As shown in Fig. \ref{Beijing}, the geometry of buildings, roads, parks, and other structures can be mapped by using geographic information system (GIS) data.\footnote{In this paper, we directly utilized the Beijing map provided by the Wireless Insite, available on: https://www.renkangtech.com/wirelessinsite}

\item \textbf{Material properties:} The materials used in the construction of buildings (e.g., glass, concrete, metal) significantly affect signal propagation. These materials' properties are included in the model to ensure realistic interactions between electromagnetic waves and the environment.

\end{itemize}

\subsection{BS and UE setup}
Once the environmental scenario is constructed, the next step is to set up the BS and UE within this environment. 
\begin{itemize}
\item \textbf{BS placement:} In an urban setting like Beijing, BSs are strategically placed to ensure coverage and capacity.
In this paper, as shown in Fig. \ref{Beijing}, 42 BSs are placed in the scenario, BS1-BS42, covering the entire map.

\item \textbf{UE location distribution:} UEs are distributed across the environment to represent typical user locations. In this paper, as shown in Fig. \ref{Beijing}, 491,040 UE locations are evenly distributed across the entire map at 2-meter intervals.

\item \textbf{BS and UE settings:} In the ray-tracing simulation, each BS and UE have only a single half-wave dipole with the axis of the dipole antenna in the z-direction and use 28 GHz mmWave for communication.
Note that output of this ray-tracing simulation can be used to generate channels for larger antenna arrays at the BSs and UEs, which will be illustrated in Section IV.
\end{itemize}

\subsection{Channel knowledge data generation}
With the scenario and setups in place, the ray tracing process simulates the propagation of electromagnetic waves from BS to UE.
\begin{itemize}
\item \textbf{Simulation parameters:} The urban complexity of Beijing requires high-resolution simulations to accurately capture reflections and diffractions. In the Wireless InSite ray-tracing simulator, the paths are configured to undergo a maximum of 6 reflections and 1 diffraction.

\item \textbf{Ray launching and propagation:} Rays are launched from the BS and traced through the environment. The interaction of rays with buildings and other structures is meticulously modeled to capture the multiple paths signals. For every transmitter-receiver pair, ray-tracing simulation shoots hundreds of rays in all directions from the transmitter and record the strongest 25 paths from those that made their ways to the receiver, where strongest paths are those paths with the highest receive power.

\item \textbf{Data collection:} The ray tracing process collects data on various channel parameters, including path loss, delay spread, AoA, and AoD. This data is recorded for each BS-UE pair, creating a comprehensive dataset that reflects the channel characteristics in Beijing's urban environment.
In this scenario, considering the coverage range of each BS to be 400$ \times $400 meters, about 1.68 million data entries are generated and stored.

\end{itemize}

%
%
%

\subsection{CKM generation}
To provide an intuitive understanding of the channel characteristics, the CKMImageNet includes visual representations in the form of channel heatmaps and physical environment maps.

\begin{itemize}
\item \textbf{CKM generation:} Heatmaps are created to visually represent the spatial distribution of key channel parameters, such as signal strength, path loss, and interference levels. The generation process involves:
1) Visualization: Using advanced algorithms to convert the normalized data into a visual format. These algorithms ensure that the heatmaps accurately depict variations in signal quality and other channel parameters.
2) Color mapping: Applying color gradients to represent different values, making it easy to identify areas with strong signals, high interference, or significant path loss at a glance.
3) Resolution and detail: Ensuring high spatial resolution to capture fine-grained variations in the channel characteristics, providing detailed insights that aid in network planning and optimization.

In this scenario, about 6,000 heatmaps are generated, each has 64$ \times $64 pixels, comprising 4,096 pixels corresponding to channel knowledge at a size of 128$\times$128 meters in physical map.

\item \textbf{Environment map generation:} Physical environment maps provide a visual representation of the actual environment in which the channels exist, including buildings, terrain, and other structures. The generation process involves:
1) Visualization: Based on the digital map generated in ray tracing software, specialized tools are used to generate and display the physical environment maps. 
2) Contextual information: Adding metadata and annotations to the maps to provide additional context, such as the locations of BSs, which helps in understanding the overall network setup and performance.

In this scenario, the selected region of the Beijing map is divided into about 2,000 smaller maps, each corresponding to a CKM.

\end{itemize}

In summary, constructing the CKMImageNet for an urban environment like Beijing involves meticulous scenario construction, strategic BS and UE setup, ray tracing simulations, and the creation of detailed channel data, visual CKM and physical environment map. This comprehensive approach ensures that the CKMImageNet accurately reflects real-world conditions, providing valuable insights for CKM construction.

\section{Applications of Channel Knowledge dataset}
CKMImageNet serves as a powerful tool with numerous applications in wireless communication. By providing location-specific channel knowledge data and image, the CKMImageNet facilitates CKM construction and communication algorithm performance verification. This section explores several key applications of the CKMImageNet, illustrating its versatility and impact on the field.

\subsection{CKM construction algorithm }

CKMImageNet enables the construction of CKM by providing detailed and location-specific channel data integrated with environment context. 
By combining numerical data with visual representations such as heatmaps and physical environment maps, CKM construction becomes more intuitive and insightful, enabling researchers to use image-based algorithm to identify and understand complex spatial relationships and intrinsic patterns in channel behavior. 
Besides, the comprehensive data in CKMImageNet supports the development of sophisticated inference algorithms that can predict channel behavior under various conditions, improving the overall effectiveness of environment-aware communication. 

\subsection{Communication algorithm performance verification }
The CKMImageNet is an invaluable resource for validating the performance of communication algorithms. By offering detailed and realistic channel data, CKMImageNet allows researchers to rigorously test new algorithms and compare their performance against established benchmarks. 
One of the standout features of the CKMImageNet is its ability to generate customized channels based on specific user requirements. This capability allows users to tailor the channel data to match their particular scenarios, offering a high degree of flexibility and precision in network design and analysis.
Specifically, consider a mmWave massive MIMO communication system, which consists of a BS with $ M_t=M_t^yM_t^z $  transmit antennas and a UE with $ M_r=M_r^yM_r^z $ receive antennas, with $  y, z $ denoting the number of antennas in $ y,z $ directions. Then, the channel $ \mathbf{H}\in\mathbb{C}^{M_r\times M_t} $can be expressed based on the channel data in CKMImageNet as
\begin{equation}
	\mathbf{H}=\sqrt{M_tM_r}\sum_{l=1}^{L}\alpha_l\mathbf{a}_r(\theta_l^e, \theta_l^a)\mathbf{a}_t^H(\phi_l^e, \phi_l^a),
\end{equation}
where $ \mathbf{a}_t(.,.) $ and $ \mathbf{a}_r(.,.) $ are the array response of the BS and UE, respectively.
The ability to customize channel scenarios ensures that algorithms are evaluated under diverse and realistic conditions, providing robust performance assessments. Researchers can configure the transmitter and receiver settings to simulate various network factors. This thorough validation process is crucial for developing reliable and high-performance communication systems.

\subsection{AI training dataset}
CKMImageNet can serve as an exceptional resource for training AI models, particularly for image-based applications. The detailed and extensive channel data provided by the CKMImageNet, including numerical information and visual components such as CKM and physical environment maps, offer a rich and diverse dataset essential for developing sophisticated AI algorithms. By leveraging this comprehensive dataset, AI models can be trained to accurately recognize patterns, predict channel behavior, and optimize network performance in real-world scenarios. The CKMImageNet's high-resolution and realistic data improve the generalization capabilities of AI models, making them more effective in diverse environments. This facilitates the advancement of AI-driven applications in wireless communication, such as automated network optimization, predictive maintenance and CKM construction, ultimately enhancing the efficiency and reliability of wireless networks.

\section{Conclusion}
CKMImageNet is a comprehensive dataset that includes both numerical channel data and visual images, which enables efficient CKM construction via CV.
By providing detailed location-based channel data and environmental context, CKMImageNet supports the verification of communication and sensing algorithms and facilitates CKM construction with AI.
Future work on will focus on expanding its capabilities by incorporating additional scenarios, frequencies, and real-world data.
CKMImageNet will ensure its pivotal role in providing the data foundation for 6G environment-aware and AI-enabled intelligent communication.



\bibliographystyle{IEEEtran}
\bibliography{IEEEfull,IEEEabrv}
	
\end{document}